\documentclass[aps,pra,twocolumn,groupedaddress,amssymb,amsmath,amsthm]{revtex4-1}
\usepackage{amsfonts}

\usepackage{amssymb}

\usepackage{verbatim}
\usepackage{amsmath}
\usepackage{graphicx}

\newcommand{\bfk}{{\bf k}}

\newcommand{\bfq}{{\bf q}}

\newcommand{\ups}{\uparrow}
\newcommand{\dos}{\downarrow}

\begin{document}
\title{Thermodynamics of Attractive and Repulsive Fermi Gases in Two Dimensions}
\author{Weiran Li} 
\affiliation{Condensed Matter Theory Center and Joint Quantum Institute, Department of Physics, University of Maryland, College Park, Maryland 20742, USA 
}
\date{\today}
\begin{abstract}

We study the attractive and repulsive two-component Fermi gas with spin imbalance in two dimensions. Using a generalized $T$-matrix approximation, we examine the thermodynamic properties of both attractive and repulsive contact interacting Fermi gases. The interaction strength, which is characterized by the bound state energy $E_b=\hbar^2/m a_{2d}^2$ in vacuum, can be adjusted through a Feshbach resonance. We calculate the interaction energy, compressibility and spin susceptibility of the two branches of the Fermi gas. For the repulsive branch, we also find a critical strength of interaction $a_{2d}^{(c)}$ above which this metastable thermodynamic state becomes unstable. This critical value depends on the temperature and the spin imbalance (the ``magnetization") of the system. 

\end{abstract}
\maketitle
\section{Introduction}

Dimensionality plays an important role in quantum problems. A well-known example is the absence of Bose-Einstein condensation in a noninteracting two-dimensional (2D) Bose gas, due to an enhancement in the low energy density of states. Beyond this simple single-particle effect, quantum many-body physics in interacting systems can also be affected by dimensionality in a more non-trivial manner. For example, interesting physics of universal scalings and hidden symmetries appear in two-dimensional quantum gases \cite{unihydro,hiddenf}. Another area that attracts more attention is the strongly interacting regime of 2D Fermi systems, where the exotic pseudogap phase appears to be significantly enhanced by strong quantum fluctuations in low dimensions. This pseudogap phase in 2D has been discussed extensively in both latticed (e.g. cuprates) and continuous (e.g. BEC-BCS crossover of 2D fermionic superfluidity) systems\cite{cuprates, crossover, Randeria2D1, Randeria2D2}.

Atomic Fermi gases in two dimensions are interesting to study for the above reasons. With the advantages of easily tuning the interaction strength between neutral atoms with Feshbach resonances\cite{Chin}, ultracold atomic gases provide an systematic way of studying quantum problems from the weakly to strongly interacting regime. Also the adjustable geometry of optical trapping potentials makes it possible to observe the crossover in dimensionalities\cite{dxover}. Optical traps with pancake geometry serve as containers of (quasi-) two-dimensional gases. Recently, there have been many experiments on various properties of 2D Fermi gases in pancake traps. In these experiments, both the scale invariance\cite{scaling} and pseudogap features have been observed\cite{psgap2d, rf2d}. In more recent experiments, studies have been extended to a highly polarized 2D Fermi gas to study the polaron physics\cite{Kohl}, where both the attractive and repulsive polaron have been studied.

On the theory side, there have been extensive studies focusing on the spectral functions. For Fermi gases with equal spin populations, for instance, the spectral function as computed using a virial expansion exhibits incoherent pairing effects at temperatures higher than $T_c$ in the nondegenerate regime\cite{Parish2, Barth}. Other general pairing effects have also been studied\cite{Demler2,FischerParish}. In another scenario of extremely polarized (i.e., highly spin-imbalanced) 2D Fermi gases, i.e., the polaron limit, studies on the spectral functions find two distinctive peaks\cite{Parish1,ParishLevinsen,Demler}, corresponding to the attractive and the repulsive branches of the Fermi polaron. Besides the spectral function, thermodynamic quantities are also of great interest. Most studies on thermodynamics focus on the superfluid transition of the 2D BCS-BEC crossover\cite{gora,Torma, Duan, Giorgini}. Recently, a Luttinger-Ward approach has been used to study the equation of state of the attractive 2D Fermi gas\cite{EOS}. Another recent work discusses the spectroscopy and thermodynamic quantities on the mean-field BCS level of Fermi gases in quasi-two-dimensions, including the transverse degree of freedom of the pancake trap\cite{ParishTQ}. These studies on thermodynamics, however, all explore the attractive Fermi gas with equal spin population only.

In this paper we calculate various thermodynamic quantities of normal Fermi gases in {\it strictly} two dimensions, where the possible occupation of higher harmonic states in the transverse direction of the pancake trap is completely neglected. We address both the attractive and repulsive branches, as well as the effect of finite spin imbalance. Our approach treats the interaction effects beyond the mean-field level by including the pairing fluctuation from iterating two-particle scattering processes, i.e, we employ the $T$-matrix approximation or the Nozieres-Schmitt-Rink (NSR) approach\cite{NSR}. While the thermodynamic quantities of the attractive branch are given by the conventional $T$-matrix approach, the repulsive branch is described by a generalized theory introduced in \cite{ShenoyHo}. We present the evolution of the interaction energy, compressibility, and the spin susceptibility of both attractive and repulsive branches, as a function of interaction strength and temperature. We also observe a thermodynamic instability in the repulsive branch, which is consistent with experiments and previous calculations of the spectral function\cite{rf2d,Kohl,Demler,Parish1}. Our approach is applicable to systems with a large range of spin imbalance, and hence our work covers a variety of 2D Fermi gases with two species from equal population to the extremely polarized case, i.e., the polaron limit.

The paper is structured as follows: in section \ref{sec:model}, we describe the generalized $T$-matrix  approximation in detail. We then show our main results for thermodynamic quantities of two-dimensional Fermi gases with equal population and with spin imbalance in \ref{subsec:equal} and \ref{subsec:imba}, respectively. Finally, we conclude and discuss our findings in section \ref{sec:conclusion}.

\section{general model} \label{sec:model}
We consider a two-component Fermi gas confined in two dimensions. Experimentally, this is realized by loading the two species of fermionic atoms (which we label as spin $\ups$ and $\dos$) to an extremely tight trap in the $z$-direction with a frequency $\omega_z$, where $\hbar \omega_z\gg E_F, k_B T$. $E_F=\pi \hbar^2 n/m$ is the Fermi energy given by the two-dimensional density $n$ of the gas, and $T$ is its temperature. The system is effectively two-dimensional in this limit, since the motion in the tight $z$-direction is frozen to the lowest harmonic oscillator state. 

From now on we set $\hbar=k_B=1$. The full Hamiltonian of the system is
\begin{equation}\label{eqn:hamiltonian}
H=\sum_{\bfk,\sigma}\epsilon_{\bfk\sigma} c^\dag_{\bfk\sigma} c_{\bfk\sigma}+\frac{g}{\Omega}\sum_{\bfk,\bfk',\bfq} c^\dag_{\bfk'\ups} c^\dag_{\bfq-\bfk'\dos} c_{\bfq-\bfk\dos} c_{\bfk\ups},
\end{equation}
where $c^\dag_{\bfk\sigma}$ is the creation operator of a fermion with momentum $\bfk$ and spin index $\sigma=\ups, \dos$, $\Omega$ is the two-dimensional volume of the gas. $\epsilon_\bfk=k^2/2m_{\sigma}$ is the kinetic energy of the fermions. The interaction is treated by the zero-range model where the coupling constant $g$ is momentum-independent. In two dimensions, a two-body bound state with binding energy $E_b\equiv (2m_{\rm red}a_{2d}^2)^{-1}$ always exists for arbitrary attractive interaction. In this paper, we only consider the equal mass situation in which the reduced mass $m_{\rm red}=m/2$ is half the fermion mass. The characteristic length scale $a_{\rm 2d}$, often being referred to as the two-dimensional scattering length, universally determines the low energy scattering amplitude $f^{-1}(k) =  -2\ln(k a_{2d})+i\pi$. This relation leads to the regularization of the two-dimensional zero-range model as $-g^{-1}=\Omega^{-1}\sum_\bfk (E_b+k^2/m)^{-1}$. Typically, the interaction strength is parametrized by the dimensionless quantity $\eta={\rm log}(k_Fa_{2d})$, where $k_F$ is the Fermi wave vector of the two dimensional gas which is related to the total density as $n=k_F^2/2\pi$.

The explicit expression of the $T$-matrix in terms of the regularized interaction parameter reads
\begin{equation}\label{eqn:Tmat}
T^{-1}({\bf q},\omega)=\frac{1}{\Omega}\sum_{\bf k}\left(\frac{\gamma({\bf q}, {\bf k})}{\omega+i0^+-\omega(q)-k^2/m}-\frac{1}{g}\right),
\end{equation}
where $\gamma({\bf q}, {\bf k})=1-n_{\uparrow}({\bf q}/2+{\bf k})-n_{\downarrow}({\bf q}/2-{\bf k})$; $n_{\uparrow}, n_{\downarrow}$ are the Fermi distribution functions for up and down spins, respectively. This $\gamma$ factor signifies the Pauli blocking. $\omega(q)\equiv q^2/4m-\mu_\ups-\mu_\dos$ is the threshold of positive energy scattering in the two-particle sector. $\mu_\ups, \mu_\dos$ are the chemical potentials of the up-spin and down-spin species, respectively. Later, we also use an alternative notation where $\mu_\ups\equiv \mu+h$ and $\mu_\ups\equiv \mu-h$, where $h$ is the effective ``polarization field", analogous to the Zeeman term for electrons in a magnetic field. 

The two-dimensional interaction parameter $\eta$ is tunable via a Feshbach resonance. The explicit relation of $\eta$ to the scattering length in three dimension (3D) is shown in \cite{a2d} :
\begin{equation}
a_{2d}\sim l_0\exp{\left(-\sqrt{\pi/2}l_0/a_s\right)} ,
\end{equation}
where $a_s$ is the scattering length in 3D, and $l_0=\hbar/\sqrt{m\omega_z}$ is the harmonic oscillator length in the direction of the tight trap. As $a_s^{-1}$ passes through the resonance from the BEC to the BCS regime, the 2D interaction parameter $\eta$ also varies in a large range of magnitude from negative to positive.

The method we apply in this work is a generalized $T$-matrix approximation, which can describe both the attractive and repulsive gas within a universal contact interacting model. This method was first introduced for 3D quantum gases in \cite{ShenoyHo}. The main point is to use different equations of state to address the attractive and repulsive branches of the gases. The underlying physics is that despite the attractive interaction in our model, there is a metastable repulsive branch in which the possible bound states are not populated. After excluding the bound state contribution to the thermodynamics, we get an effective description of the repulsive Fermi gas. 

In a general $T$-matrix approximation, the total density $n$ can be written as a sum of the free particle contribution $n_0$ and an interaction part $\Delta n$, i.e., $n=n_0+\Delta n$. Our generalized theory for the repulsive branch of the gas further separates the interaction contribution $\Delta n$ into two parts,  a scattering part $\Delta n^{sc}$, and a bound state part $\Delta n^{bd}$. The equations of state for the two branches are written as
\begin{eqnarray}\label{eqn:natt}
n_{{\rm att}}(\mu, h, T)&=& n_0+\Delta n^{sc}+\Delta n^{bd},\\
n_{{\rm rep}}(\mu, h, T)&=& n_0+\Delta n^{sc},\label{eqn:nrep}
\end{eqnarray}
where $n_0(\mu,h,T)=\Omega^{-1}\sum_\bfk \left(f(\mu+h,T,\bfk)+f(\mu-h,T,\bfk)\right)$ is the sum of the Fermi distribution functions for the two species. All other thermodynamic quantities (pressure, energy density etc.) can be derived from the above equation of state using the Gibbs-Duhem relation. 

In the $T$-matrix approximation, the equation of state only depends on the phase angle and the pole structure of the $T$-matrix. The explicit expressions for the scattering and bound state parts are
\begin{equation}
\Delta n^{sc}(\mu,h,T)=  -\frac{1}{\Omega}\sum_\mathbf{q}\int^{\infty}_{\omega(q)} \frac{\mathrm{d}\omega}{\pi}\, n_B(\omega)\frac{ \partial \zeta ({\bf q}, \omega)}{\partial \mu},
\end{equation}
\begin{equation} 
\Delta n^{bd}(\mu,h,T)= -\frac{1}{\Omega} \sum_\mathbf{q} n_{B} (\omega_{b}(q))
\frac{ \partial \omega_{b}(q)}{\partial \mu}, 
\end{equation}
where $n_B$ is the Bose distribution function, $\zeta ({\bf q}, \omega)={\rm arg}\, T^{-1}$ is the phase angle of the inverse $T$-matrix (\ref{eqn:Tmat}), and $\omega_b(q)<\omega(q)$ is the bound-state pole. In vacuum, $\omega_b(q)=\omega(q)-E_b$ always appears for any interaction strength and center-of-mass momentum. In the medium, however, it has been shown that the Fermi statistics blocks the pairing by the Pauli principle, and the bound state can be absent in some situations\cite{Pekker, ShenoyHo}. In our 2D model, when $q^2/4>2\mu$, the Pauli blocking effect is not strong enough such that bound states with center-of-mass momentum $q$ always exist\cite{SR}. 

For the spin-imbalanced Fermi gas, namely $\mu_\ups>\mu_\dos$ or $h>0$ (here we always choose $\ups$ as the majority component, since the two species are symmetric---the physics are the same for $h<0$), we have another set of equations for the ``magnetization", i.e., the difference in densities of the two species $M=n_\ups-n_\dos$. A corresponding normalized ``polarization" is defined as $m=M/n=(n_\ups-n_\dos)/(n_\ups+n_\dos)<1$. The $m\rightarrow 1$ limit is the polaron limit. The equations for the magnetization read
 \begin{eqnarray}
\Delta M^{sc}(\mu,h,T)&=&  -\frac{1}{\Omega}\sum_\mathbf{q}\int^{\infty}_{\omega(q)} \frac{\mathrm{d}\omega}{\pi}\, n_B(\omega)\frac{ \partial \zeta ({\bf q}, \omega)}{\partial h},\\
\Delta M^{bd}(\mu,h,T)&=& -\frac{1}{\Omega} \sum_\mathbf{q} n_{B} (\omega_{b}(q))
\frac{ \partial \omega_{b}(q)}{\partial h}. \label{eqn:Mbd}
\end{eqnarray}
The total magnetization of the 2D gas for the attractive and repulsive branch is similar to the number equation:
\begin{eqnarray}\label{eqn:Matt}
M_{{\rm att}}(\mu, h, T)&=& M_0+\Delta M^{sc}+\Delta M^{bd},\\
M_{{\rm rep}}(\mu, h, T)&=& M_0+\Delta M^{sc}.\label{eqn:Mrep}
\end{eqnarray}

Equations (\ref{eqn:natt})-(\ref{eqn:Mrep}) are the basis of our calculation. For a given temperature and polarization, we solve the corresponding chemical potential $\mu$ and the polarization field $h$ to further obtain the thermodynamic quantities. In our paper we limit our parameter space to $\mu<0$, which is enforced by either relatively high temperatures or high polarizations. This limitation comes from a divergence in number equation for the repulsive branch in the $T$-matrix approximation, which can be fixed by considering a fluctuation in the saddle point position\cite{beyondNSR}. 

\section{Results for thermodynamic quantities}\label{sec:result}

\subsection{Equal-Spin Gases} \label{subsec:equal}

\begin{figure}
\includegraphics[width=0.45\textwidth]{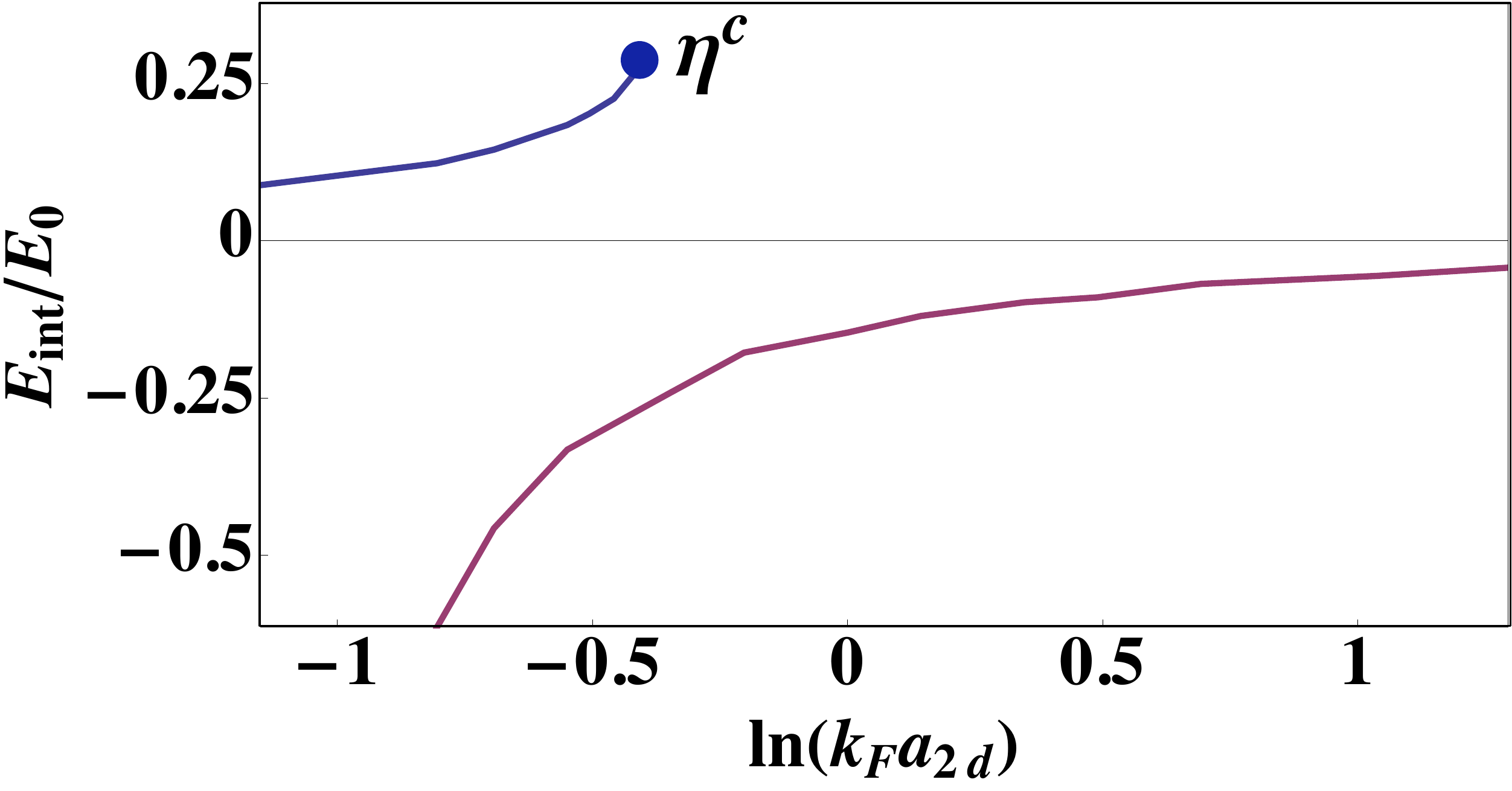}
\caption{\label{fig:E6TF}The interaction energy for the attractive branch (red curve) and repulsive branch (blue curve) at a high temperature $T=6T_F$ for a system with equal spin population. The end point of the repulsive branch is located at $\eta_c=\ln{k_F a_{2d}^c}\approx -0.4$.}
\end{figure}

We first calculate the thermodynamic quantities for equal-spin Fermi gases at relatively high temperatures, where $\mu<0, h=0$. As discussed in section \ref{sec:model}, a bound state always appears for any center-of-mass momentum. Thus, the repulsive branch is always distinct from the underlying attractive branch. Similar to the BCS-BEC crossover in 3D, the attractive branch shows an increasing attraction energy as $\eta={\rm ln} (k_F a_{2d})$ decreases, as shown in Fig. (\ref{fig:E6TF}), where the gas is in the non-degenerate regime with $T=6T_F$. This feature has been experimentally observed in rf-spectroscopy measurements\cite{rf2d}. The majority of particles evolve from shallow dimers to deeper bound diatomic molecules as $E_b$ increases. Even in the high temperature regime, we see that the interaction energy is comparable to the kinetic energy at the ``2D unitarity" $\eta=0$. In 3D, this large attractive interaction leads to a high temperature superfluid transition with a large $T_c/T_F$ ratio\cite{crossover}. In 2D, however, due to a different Berezinskii-Kosterlitz-Thouless (BKT) mechanism of superfluid transitions, $T_c/T_F$ is much lower than Fermi gases in 3D\cite{EOS}.
\begin{figure}[t]\centering
\includegraphics[width=0.4\textwidth]{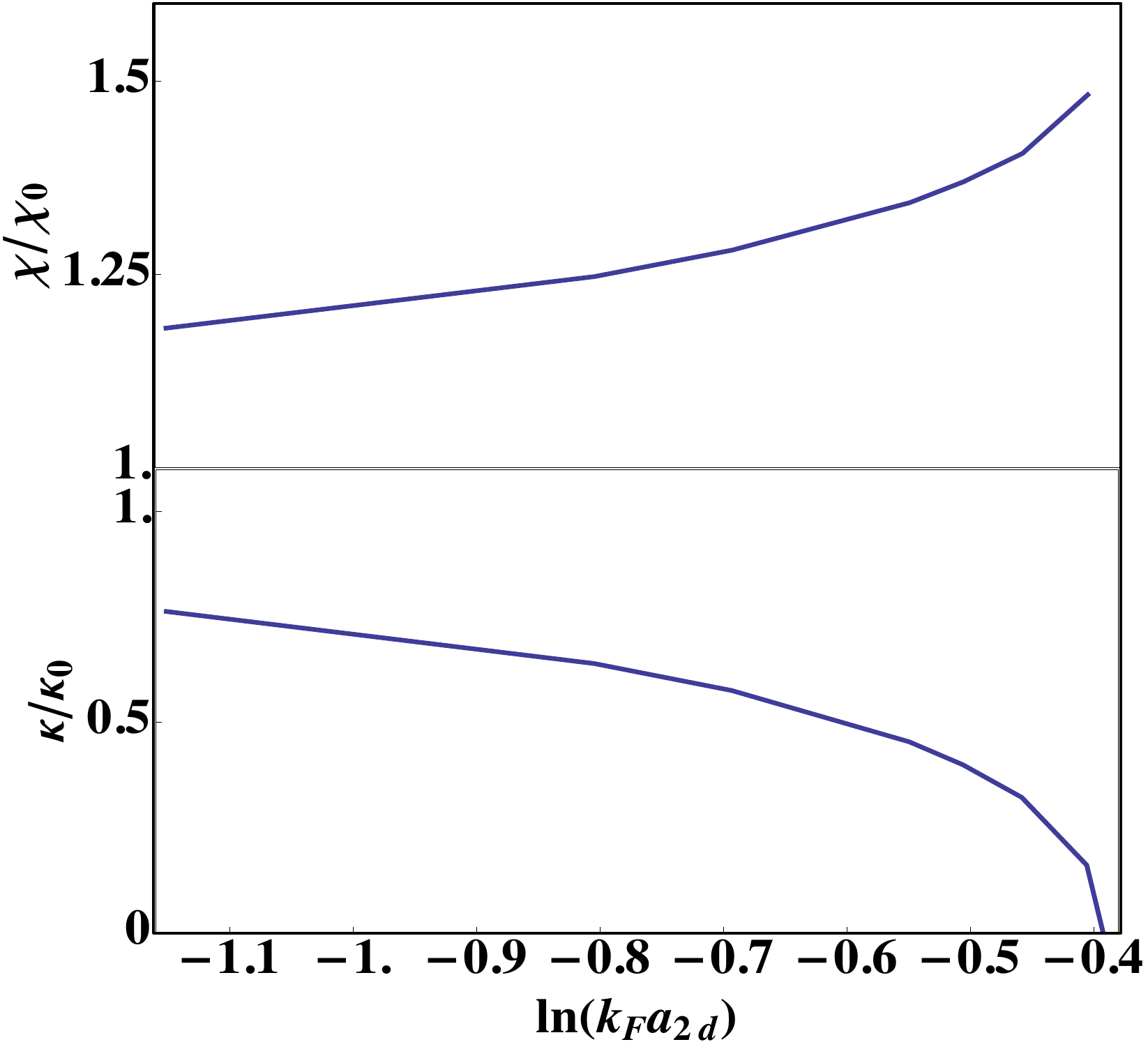}
\caption{\label{fig:chikappa2d}Spin susceptibility (upper panel) and compressibility (lower panel) of the repulsive branch at high temperature $T=6T_F$, rescaled by the noninteracting susceptibility $\chi_0$ and compressibility $\kappa_0$ at the same temperature. As the repulsive interaction increases, the system becomes more and more rigid with lower compressibility value. At $\ln{k_F a_{2d}^c}=-0.39$, it becomes ``incompressible" such that the upper branch will collapse if the repulsive interaction is further increased. This is the corresponding critical point of the stability in the upper branch. Within the stable region $\ln{k_F a_{2d}}<-0.4$, $\chi/\chi_0$ is at most 1.5, indicating the absence of a ferromagnetic transition at this temperature.}
\end{figure}

Now we focus on the repulsive branch of the 2D Fermi gas. The repulsive branch is a metastable state with finite lifetime, which is determined by the three-body collision rate. In the high temperature regime, the three-body recombination is substantially suppressed and hence the repulsive branch may persist in a well-defined thermodynamic state. We observe an increasing repulsive energy as $\eta$ increases in Fig. (\ref{fig:E6TF}), which is consistent with the observations in \cite{rf2d}. For a large repulsive interaction, the upper branch is unstable, similar to the 3D Fermi gas\cite{Pekker,ShenoyHo}. From spectroscopic observations, this is revealed by a broadening of the atomic peak, which indicates a shorter lifetime of the repulsive branch when the repulsion increases\cite{Kohl,Demler,Parish1}. From our thermodynamic perspective, the repulsive branch terminates sharply at a critical interaction strength $\eta^c={\rm ln}k_Fa_{2d}^c$, as shown in Fig. (\ref{fig:E6TF}) by the solid dot, above which the upper branch no longer stays in a well-defined state. The physics of this instability is explained in the following: when the interaction strength varies from the weakly repulsive limit, a decrease in the compressibility is predicted from perturbation theory. The compressibility further decreases with stronger repulsions, until it reaches zero at a critical value of $a_{2d}^c$, where the gas becomes unstable. This can be understood heuristically using a hard-sphere picture: when the repulsive interaction becomes larger, the system appears more and more rigid, before it finally becomes incompressible as the atoms are closely packed. This mechanism is absolutely different from conventional mechanical instability, where the compressibility diverges and the system collapses. 

Mathematically, our method relies on finding the solutions for $\mu$ and $h$ at different temperatures, polarizations, and interaction parameter $\eta$. While for the attractive branch, the solutions of $\mu$ and $h$ always exist for any interaction strength, in the repulsive branch it is not guaranteed to find these solutions. At certain temperature and interaction strength, there is a maximum of equation of state $n(\mu,T)$ as a function of $\mu$. The density maximum corresponds to a point of zero compressibility, and this point is the instability point if the maximum gives the solution to the number equation (\ref{eqn:nrep}).

We plot the compressibility and spin susceptibility of the repulsive Fermi gas within the stable region $\eta<\eta^c$ in Fig. (\ref{fig:chikappa2d}). The results are rescaled by the noninteracting values at the same temperature. As in Fig. (\ref{fig:chikappa2d}), we clearly see the monotonically decreasing compressibility and increasing spin susceptibility for increasing $\eta$. Due to enhanced fluctuations in the spin sector in 2D, $\chi$ is generally larger than the 3D situation with the same temperature. However, the largest value of $\chi/\chi_0$ is only around $1.5$ within the stable region of the repulsive branch, indicating that a ferromagnetic transition is absent at this temperature. 

\begin{figure}
\includegraphics[width=0.41\textwidth]{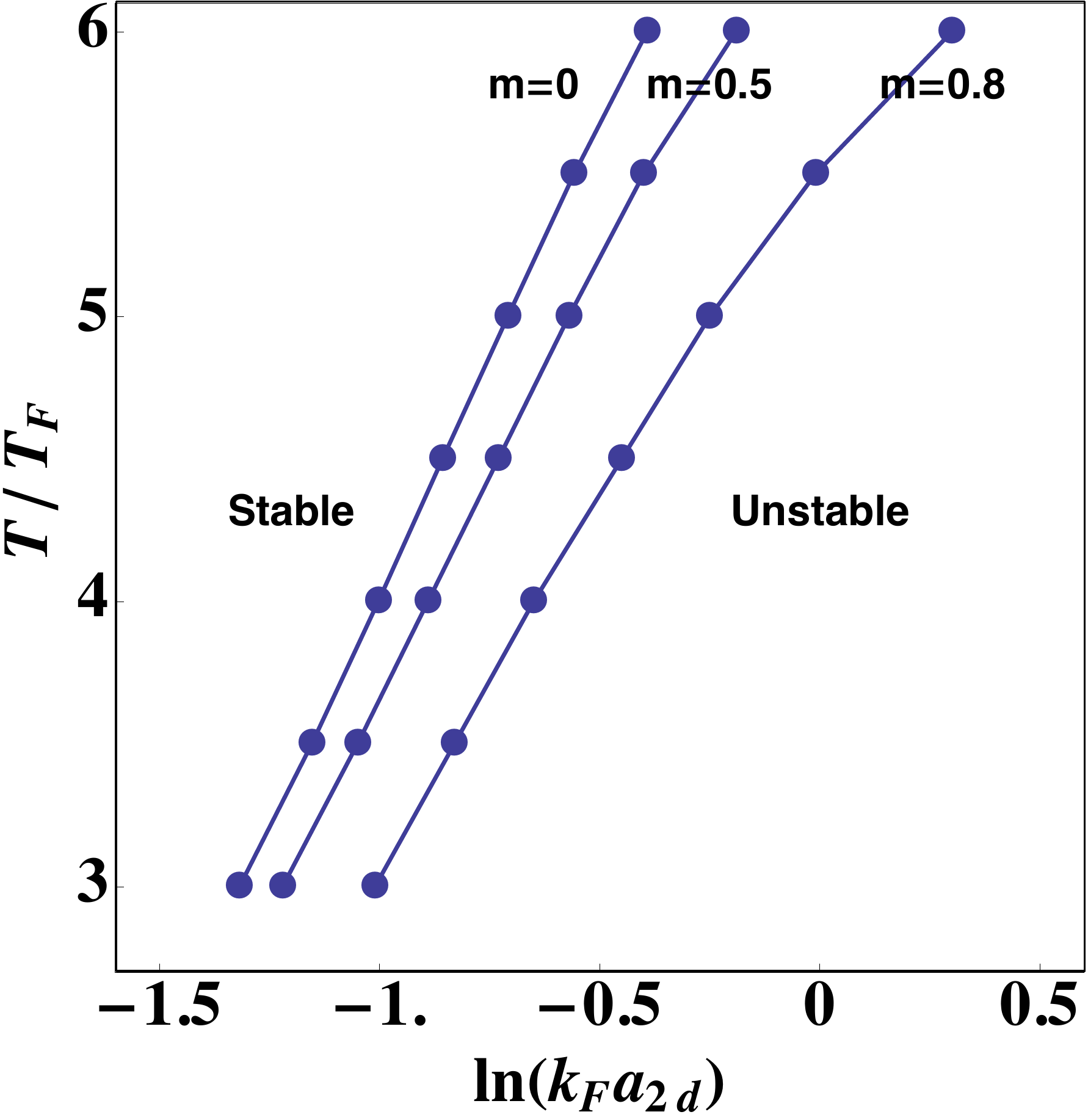}
\caption{\label{fig:stab}A stability diagram of the repulsive Fermi gas. The three different curves correspond to the $\kappa=0$ contour at different magnetizations $m=0, 0.5, 0.8$. To the left of the curves, the repulsive branch of a two dimensional gas can be stable, while it has zero compressibility at the boundaries and collapses to the right. The repulsive branch can have larger region of stability at higher temperatures or spin imbalances, all come from the suppression of repulsive interactions in the system.}
\end{figure}
\subsection{Spin-Imbalanced Gases} \label{subsec:imba}

A spin imbalance in the 2D Fermi gases can easily be included in our approach. Recent experiments have successfully created the extremely polarized 2D Fermi gas in the polaron limit\cite{Kohl}. Later theoretical works on the repulsive polarons all find the instability in the spectroscopy. Here, we determine the stability boundary as a function of polarization. Our result goes to the polaron limit for $m\rightarrow 1$.  

The difference in spin population is induced by setting the polarization field $h\neq 0$. The polarization is adjusted by $h$ and the polaron limit is set by $h\rightarrow \infty$. In our current paper we only take the largest polarization to $m=0.8$. This number is also close to the polarization limit that can be reached in realistic experiments. For two-component fermions, since the $s$-wave contact interaction only appears in the interspecies channel, the interaction effect is suppressed when the polarization increases. Consequently, the stable region of the repulsive Fermi gas is pushed further into the stronger repulsive (large $\eta$) region. Fig. (\ref{fig:stab}) shows the stability boundaries for different polarizations at various temperatures. We can clearly see that the boundary moves to larger $\eta$ for finite polarizations. Also, for higher temperatures, as the interaction effect is suppressed, the boundary naturally moves to the stronger repulsive direction. Basically, any effect that suppresses the repulsive energy will enhance the stability of the upper branch of the Fermi gas.

\section{Conclusions and Discussions}\label{sec:conclusion}

In this paper we discuss the interaction energy, compressibility and spin susceptibility of attractive and repulsive two-dimensional contact-interacting Fermi gases with spin imbalance. We apply a generalized $T$-matrix approximation to calculate the thermodynamics of the attractive and repulsive branches. The attractive branch evolves continuously from the BCS limit to the BEC limit with increasing interaction strength. For the repulsive branch, on the other hand, an instability is observed where the compressibility vanishes. This thermodynamic instability is consistent with recent experiments and theoretical calculations of the spectral functions\cite{rf2d,Kohl,Demler,Parish1}, and we interpret it heuristically by a hard sphere picture. Our work also provides a universal theory to study the thermodynamics of the 2D Fermi gas with any spin imbalance, from the equal-population case to the polaron limit. 

The approach we apply in this paper is limited to a relatively high temperature regime where the chemical potential is negative. This is due to the divergence of the number equation at positive chemical potentials. This only happens for dimensions lower than three. This divergence can be fixed by introducing a fluctuation of the saddle point position\cite{beyondNSR}, which is beyond the non-self-consistent or the self-consistent $T$-matrix approximation, and is saved for future studies. Also a straightforward extension of our work, the polaron limit, i.e., close to unity magnetization $m=1$ at low temperatures is also an open question.

\section{Acknowledgments}
The author gratefully acknowledges discussions with Tin-Lun Ho on the scattering amplitude of the 2D models. This work is supported by NSF-JQI-PFC, ARO-Atomtronics-MURI, and partly by DARPA-OLE.


\end{document}